\documentclass[prb,aps,twocolumn,showpacs,superscriptaddress,amsmath,amssymb]{revtex4}

\usepackage{graphicx}
\usepackage{dcolumn}
\usepackage{bm}

\begin{document}

\title{      Origin of charge density wave in the coupled spin 
             ladders in Sr$_{14-x}$Ca$_x$Cu$_{24}$O$_{41}$ }

\author{     Krzysztof Wohlfeld}
\affiliation{Marian Smoluchowski Institute of Physics, Jagellonian
             University, Reymonta 4, PL-30059 Krak\'ow, Poland }
\affiliation{Max-Planck-Institut f\"ur Festk\"orperforschung,
             Heisenbergstrasse 1, D-70569 Stuttgart, Germany }

\author{     Andrzej M. Ole\'s}
\affiliation{Marian Smoluchowski Institute of Physics, Jagellonian
             University, Reymonta 4, PL-30059 Krak\'ow, Poland }
\affiliation{Max-Planck-Institut f\"ur Festk\"orperforschung,
             Heisenbergstrasse 1, D-70569 Stuttgart, Germany }

\author{     George A. Sawatzky}
\affiliation{Department of Physics and Astronomy, University of 
             British Columbia, Vancouver B. C. V6T-1Z1, Canada }

\date{\today}

\begin{abstract}
We formulate a $d$-$p$ multiband charge transfer model for Cu$_2$O$_5$ 
coupled spin ladders, relevant for Cu$_2$O$_3$ plane of 
Sr$_{14-x}$Ca$_x$Cu$_{24}$O$_{41}$, and solve it using Hartree-Fock 
approximation. The results explain that 
(i) the charge density wave (CDW) with its periodicity dependent 
on doping is stabilized by purely electronic many-body interactions 
in {\it a single spin ladder} and
(ii) the inclusion of the interladder interactions favors (disfavors) 
the stability of the CDW with odd (even) periodicity, respectively.
This stays in agreement with recent experimental results and suggests
the structure of the minimal microscopic model which should be 
considered in future more sophisticated studies.
\end{abstract}

			     
\maketitle

\begin{figure}[b!]
\includegraphics[width=8cm]{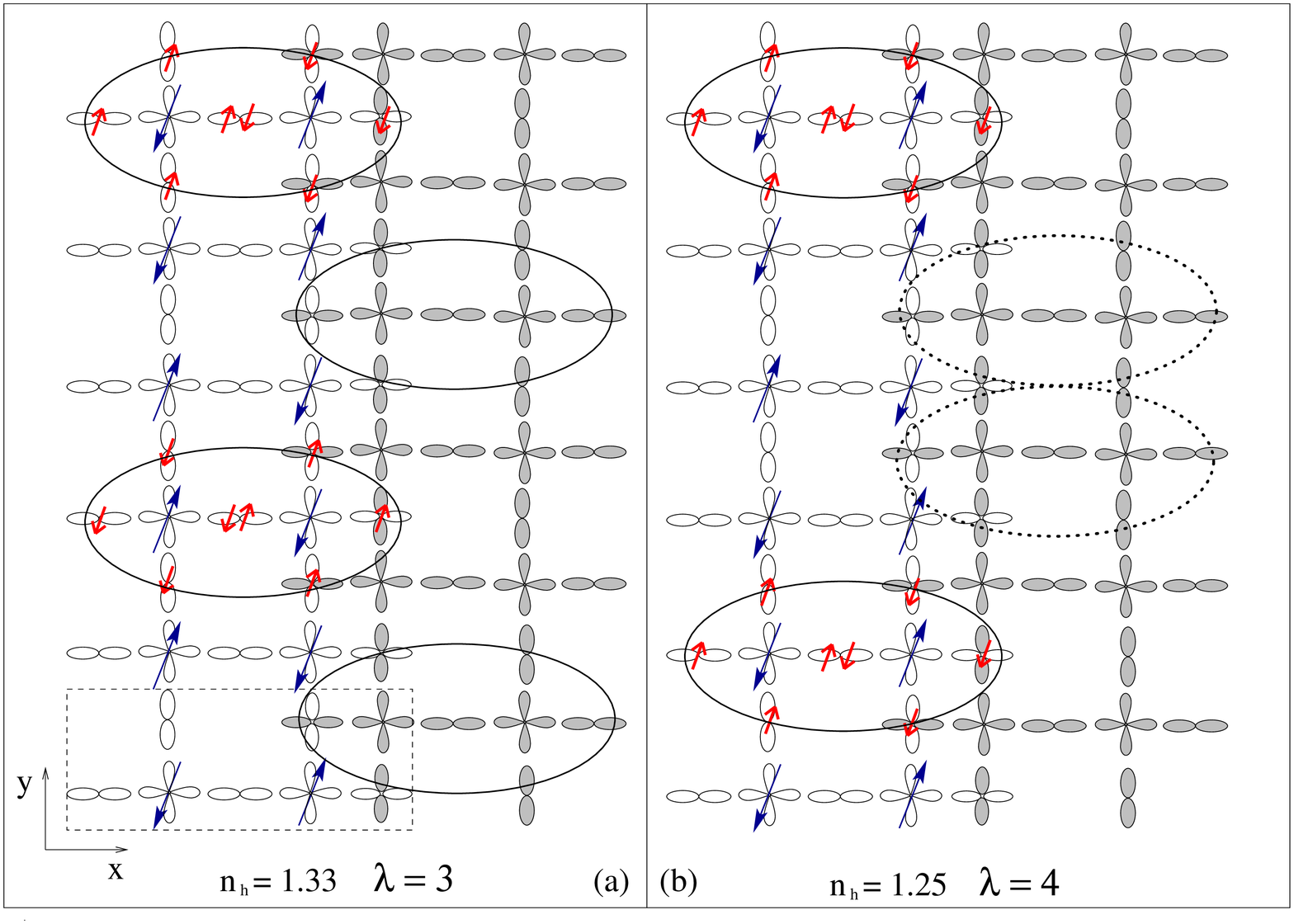}
\caption{(Color online) 
Schematic picture of two coupled Cu$_2$O$_5$ ladders (white and grey)
with a CDW order of period: 
(a) $\lambda=3$ and 
(b) $\lambda=4$. 
The Cu$_2$O$_5$ unit cell with two $3d_{x^2-y^2}$, three $2p_x$, 
and two $2p_y$ orbitals is indicated by dashed line.
The arrows stand for hole spins in Cu and O orbitals, with their 
(large) small size corresponding to +1.0 (+0.25) hole charge. 
The ovals show rungs with enhanced hole density in the CDW phase.
The dotted ovals in the grey ladder of (b) show 
the two possible degenerate states, see text. 
}
\label{fig:art}
\end{figure}

Sr$_{14-x}$Ca$_x$Cu$_{24}$O$_{41}$ (SCCO) is a layered material with 
two distinctly different types of two-dimensional (2D) copper oxide 
planes separated by Sr/Ca atoms:\cite{Blu02} 
(i) the planes with almost decoupled CuO$_2$ chains and 
(ii) the Cu$_2$O$_3$ planes formed by Cu$_2$O$_5$ coupled ladders 
(see Fig. \ref{fig:art}). 
The latter ones exhibit the non-BCS superconducting (SC) phase for 
$x=13.6$ in SCCO under pressure larger than 3 GPa,\cite{Ueh96} or a 
spin-gapped insulating charge density wave (CDW) state in broad range of 
$x$ and under normal pressure.\cite{Vul05} By means of the resonant soft 
x-ray scattering it was found\cite{Abb04} that this CDW is driven by 
many-body interactions (presumably just Coulomb on-site interactions 
since the long-range interactions are screened in copper oxides
\cite{Gra92}), and it cannot be explained by a conventional 
Peierls mechanism. Hence, the observed competition between the CDW 
(also referred to as the "hole crystal" due to its electronic origin) 
and SC states in spin ladders resembles the one between stripes and the
SC phase in CuO$_2$ planes of a high $T_c$ superconductor (HTS), which 
makes the problem of the origin of the CDW phase in SCCO both generic 
and of general interest.

Furthermore, recently it has been found\cite{Rus06a} that 
the only stable CDW states are with period $\lambda=5$ for $x=0$,
and with period $\lambda=3$ for $x=11$ 
(and with a much smaller intensity for $x=10$ and $12$), 
while such a CDW order has not been observed for $1\leq x\leq 5$. 
These striking results, which contradict the previous suggestion 
\cite{Vul05} that the CDW order occurs in the entire range of 
$0\leq x < 10$, need to be explained by considering 
hole density per Cu site increasing with $x$. Recently\cite{Rus06b} 
a much higher hole density in the ladders was reported than believed 
before,\cite{Nue00} i.e., $n_h=1.20$ (number of holes/Cu ion) for 
$x=0$, $n_h=1.24$ for $x=4$, and $n_h=1.31$ for $x=11$. The aim of this 
paper is to explain theoretically these puzzling features of the CDW order 
using the above hole density.

On the one hand, it is widely believed\cite{Not07} that a two-leg spin 
ladder described by the $t$--$J$ model captures the essential physical 
properties of SCCO. The idea that merely on-site Coulomb interactions 
could lead to charge ordering was already suggested by White {\it et al.\/}
\cite{Whi02} using density matrix renormalization group (DMRG) 
--- they found that a CDW of period $\lambda=4$ is the 
(possibly spin gapped) ground state at $n_h=1.25$. 
It is, however, quite remarkable that 
{\it such\/} a CDW has not been observed.\cite{Rus06a} 
On the other hand, the validity of the $t$-$J$ model for 
Cu$_2$O$_5$ coupled spin ladders is not obvious since: 
(i) unlike the CuO$_2$ plane of a HTS, a single Cu$_2$O$_5$ ladder lacks 
the $D_{4h}$ symmetry making the Zhang-Rice (ZR) derivation\cite{Zha88} 
of the $t$-$J$ model questionable and
(ii) Cu$_2$O$_5$ spin ladders are coupled through the on-site Coulomb 
interactions between holes in different O($2p$) orbitals, so new 
interactions could arise.

This suggests that the multiband charge transfer model\cite{Zaa85} 
adapted to the Cu$_2$O$_5$ ladder geometry, similar to those introduced
earlier for CuO$_2$ planes\cite{Ole87} or CuO$_3$ chains\cite{Ole91} of HTSs, 
could be more appropriate to capture the essential physical phenomena.
As parameters the charge transfer model includes: 
the energy for oxygen $2p$ orbital $\Delta$, 
the $d$-$p$ hopping $t$ between the nearest neighbor Cu and O sites, 
and the on-site Coulomb repulsion $U$ ($U_p$) on the Cu (O) sites, 
respectively. By solving this model in the Hartree-Fock (HF) 
approximation, we investigate whether: 
(i) the Coulomb on-site repulsion stabilizes the observed CDW 
of the respective period $\lambda$ for a given number of holes $n_h$, 
(ii) the different stability of the CDW order with odd/even period follows, 
and
(iii) the ZR singlets are formed in the spin ladder geometry. 

We consider the charge transfer model in hole notation
\begin{align}
\mathcal{H}&= \Delta\Big(\!\sum_{j\in R,L; \alpha}n_{j \alpha} 
+ \varepsilon\sum_{l}n_{lb}\Big)+\!
\Big\{\!\sum_{m,j\in R,L;\sigma}\!\!\!t_{mj} d^\dag_{m\sigma}x_{j\sigma}^{} 
\nonumber \\
&+\!\!\sum_{m,j\in R,L;\sigma}\!\!t_{mj} d^\dag_{m\sigma}y_{j\sigma}^{}\!
+\!\!\sum_{m\in R,L; l\sigma}\!\!t_{ml} d^\dag_{m\sigma}b_{l\sigma}^{}
+\mbox{H.c.}\Big\} \nonumber \\
&+U\sum_{m\in R,L} n_{md\uparrow} n_{md\downarrow}
 +U_p\!\sum_{j\in R,L; \alpha}n_{j \alpha \uparrow}n_{j \alpha \downarrow} 
\nonumber \\
&+U_p\sum_{l}n_{lb\uparrow}n_{lb\downarrow}+
 U_p\!\sum_{j\in R,L;\sigma}
 \!\Big\{(1-2\eta)(n_{jx\sigma}\tilde{n}_{jy\bar{\sigma}} 
\nonumber \\
&+n_{jy\sigma}\tilde{n}_{jx\bar{\sigma}})
+(1-3\eta)(n_{jx\sigma}\tilde{n}_{jy\sigma} 
+n_{jy\sigma}\tilde{n}_{jx\sigma})\Big\},  
\label{eq:one}
\end{align}
where $|t_{mj}|=|t_{ml}|=t$, $\alpha=x,y$, and $\bar{\sigma}=-\sigma$ 
for $\sigma=\uparrow,\downarrow$. 
The parameter $\eta=J_H/U_p=0.2$ stands for a realistic value of Hund's 
exchange ($U_p$ is the intraorbital repulsion),\cite{Gra92} and 
$\varepsilon=0.92$ yields the correct orbital energy at bridge positions.
\cite{Mue98} The model of Eq. (\ref{eq:one}) includes seven orbitals per 
Cu$_2$O$_5$ ladder unit cell (see Fig. \ref{fig:art}): 
two Cu($3d_{x^2-y^2}\equiv d$) orbitals on the right or left ($R$ or $L$) 
leg, two O($2p_y\equiv y$) orbitals on the $R/L$ leg,
two O($2p_x\equiv x$) side orbitals on the $R/L$ leg, and
one O($2p_x\equiv b$) bridge orbital on the rung of the ladder. 
We emphasize that the last two terms in Eq. (\ref{eq:one}) account for 
interladder interaction -- the holes within two different orbitals on
a given oxygen ion in a leg belong to two neighboring ladders (shown as 
white/grey orbitals in Fig. \ref{fig:art}), and are described by
charge operators $n_{jx(y)\sigma}$ with/without tilde sign in Eq. 
(\ref{eq:one}). This makes the model Eq. (\ref{eq:one}) implicitly 2D, 
though the band structure is one-dimensional (1D) when the interoxygen 
hopping $t_{pp'}$ is neglected (in fact,\cite{Gra92} $t_{pp'}\ll t$).
{\it A priori\/}, $\tilde{n}_{j\alpha}$ should be treated as particle number 
operators belonging to the Hilbert subspace of the neighboring ladder, 
resulting in a 2D many-body problem. Here we simplify it and treat 
$\rho_{j\alpha}\equiv\langle\tilde{n}_{j\alpha}\rangle$ as "external" 
classical fields adjusted to the symmetry of the CDW state, which are 
self-consistently determined within the HF approximation. 

We have solved the Hamiltonian (\ref{eq:one}) for various values of 
the model parameters $\{U,\Delta,U_p\}$, and for three different hole
densities\cite{Rus06b} $n_h=1.20,1.25,1.33$ (which correspond to the actual 
filling in SCCO in the range of $0\leq x\leq 11$) using HF approximation, 
i.e., we decouple 
$n_{j\mu\uparrow}n_{j\mu\downarrow}\rightarrow 
 \langle n_{j\mu\uparrow}\rangle n_{j\mu\downarrow}
+n_{j\mu\uparrow} \langle n_{j\mu\downarrow}\rangle
-\langle n_{j\mu\uparrow}\rangle \langle n_{j\mu\downarrow}\rangle$, 
where $\mu=d,x,y,b$. The ground state was found by diagonalizing 
the resulting one-particle Hamiltonian in real space for a single ladder 
with 60 unit cells, separately for spin up and spin down. 
The classical fields $\{\rho_{j\alpha}\}$ and $\{\langle n_{j\mu\sigma}\rangle\}$ 
were determined self-consistently with the initial values for these 
fields suggested by recent experiment,\cite{Rus06b} 
see Fig. \ref{fig:art}. While a uniform spin density wave (SDW) is 
stable for $n_h=1.0$, one finds a CDW superimposed on the SDW order for 
realistic hole densities $n_h\ge 1.20$. The stability of this composite
order follows from the 1D polaronic defects in the SDW state.
We limit the present analysis to the stability of this particular CDW 
phase, while we do not study here the possible competition with other 
phases.\cite{Woh07}

For each state we evaluate the CDW order parameter
\begin{align}
p&\!\equiv\!\sum_{i\in rung}\!\langle n_{id}+n_{ib}+n_{ix}\rangle\! 
-\!\frac{1}{\lambda -1}\sum_{i\notin rung}\!
\langle n_{id}+n_{ib}+n_{ix}\rangle 
\nonumber\\
&+\sum_{i\in rung}\langle n_{i y}\rangle
-\frac{2}{\lambda -2}\sum_{i\notin rung}\langle n_{iy}\rangle,
\label{eq:p}
\end{align}
where $\lambda$ is the period of the CDW state, and the ZR "dispersion"
defined with respect to the hole density distribution for an "idealized" 
ZR singlet state ($n_0=0.25$),
\begin{equation}
\sigma^2\!\equiv\!\!\sum_{i\in rung}\!\!
\Big\{(\langle n_{ib}\rangle - 2n_0)^2\! 
+\!(\langle n_{ix} \rangle - n_0)^2
\!+\!(\langle n_{iy} \rangle - n_0)^2\Big\}.
\label{eq:sigma}
\end{equation} 
Here and in what follows by "rung" we mean the "rung with enhanced 
hole density" which consists of seven O (four $y$, two $x$ and one $b$) 
orbitals and two Cu orbitals (see the ovals in Fig. \ref{fig:art}). Hence, 
in both above definitions the mean values of the particle number 
operators are calculated for these rungs ($i\in rung$) or for all 
remaining sites ($i\notin rung$). 
Note that in the ideal CDW phase (shown in Fig. \ref{fig:art}) $p=2$ 
and $\sigma^2=0$, irrespectively of the actual period $\lambda$.  
We also introduce {\em rung\/} hole densities on O and Cu sites
\begin{equation}
\label{eq:n}
n_{p} \!\equiv\!\sum_{i\in rung}\!\langle n_{ib}+n_{ix}+n_{iy}\rangle, 
\quad
n_{d}\!\equiv\!\sum_{i\in rung}\!\langle n_{id} \rangle. 
\end{equation} 
Similarly, magnetic order parameters are
\begin{eqnarray}
\label{eq:mo}
m_{p} \equiv&&\!\!\!\Big|\!\sum_{i\in rung\cap L}\!\!m_{ix}\!
+\!m_{iy}\Big|+\Big|\!\sum_{i\in rung\cap R}\!\!m_{ix}\!+\!m_{iy}\Big|, \\
\label{eq:mc}
m_{d}\!\equiv&&\!\sum_{i\in rung}|m_{id}|, 
\end{eqnarray} 
where the magnetization for orbital $\mu$ at site $i$ is
$m_{i\mu}=\langle n_{i\mu\uparrow}-n_{i\mu\downarrow}\rangle$. 
We recall that when holes on the rungs form two localized ZR singlets
next to each other, then $n_{d}=m_{d}\simeq 2$, $n_{p}\simeq 2$, and 
$m_{p}\simeq 1.5$, see Fig. \ref{fig:art}. 

\begin{figure}[t!]
\includegraphics[width=8.2cm]{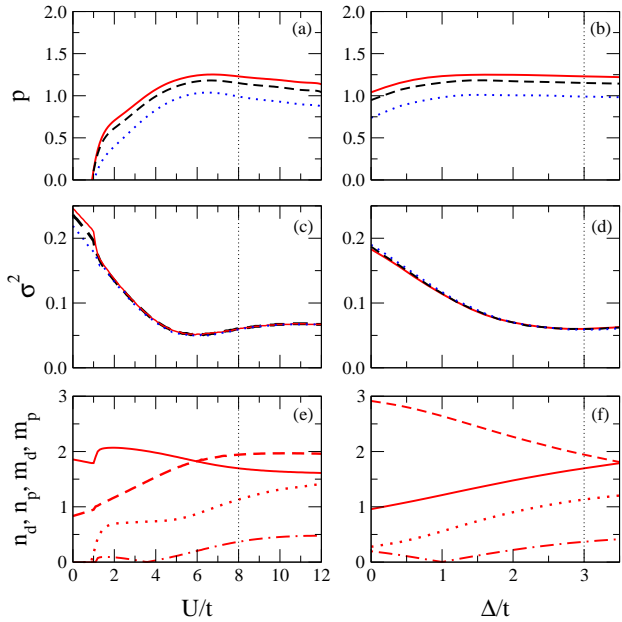}
\caption{(Color online) 
Characterization of the CDW ground states obtained with $U_p=0$ for 
increasing $U$ (left, $\Delta=3t$) and $\Delta$ (right, $U=8t$):
(a), (b) CDW order parameter $p$, and 
(c), (d) ZR singlet dispersion $\sigma^2$, for 
$\lambda=5,4,3$ shown by solid, dashed, and dotted lines, 
respectively;
(e), (f) for $\lambda=5$ charge (magnetization) in the rung 
on Cu sites shown by solid (dotted) line and on O sites  
shown by dashed (dashed-dotted) line, see Eqs. 
(\ref{eq:n})--(\ref{eq:mc}). The realistic 
values (Ref. \onlinecite{Nish02}) of $U=8t$ and $\Delta=3t$ are 
marked by vertical lines. 
}
\label{fig:den}
\end{figure}

First, we investigate the onset of the CDW phase in a single ladder of 
Fig. \ref{fig:art} by assuming $U_p=0$. In the charge transfer regime 
(for $\Delta=3t$ following Ref. \onlinecite{Nish02}) the CDW is stable 
already for $U \geq t$ with periods: $\lambda=5$ for $n_h=1.20$, 
$\lambda=4$ for $n_h=1.25$, and $\lambda=3$ for $n_h=1.33$ [Fig. 
\ref{fig:den}(a)]. For higher values of the on-site Coulomb repulsion 
$U$, $p$ first increases quite fast irrespectively of the actual CDW 
period, and next saturates at $p\sim 1$, being only about 50\% of the 
maximal value $p=2$ (a weak decrease of $p$ for $U>6t$ follows from the
charge redistribution). In particular, such a CDW order is robust for the 
widely accepted value of $U=8t$ for copper oxide ladders.\cite{Nish02} 

In the strong coupling regime of $U>4t$ the CDW is formed by holes 
distributed as in the ZR singlets since then $\sigma^2 \sim 0.05$ is 
indeed very small for all periods [Fig. \ref{fig:den}(c)]. This is also 
visible in Fig. \ref{fig:den}(e) where, in this regime, both the number 
of holes on O sites ($n_{p}$) and on Cu sites ($n_{d}$) in the rungs 
are not far from their values in the localized ZR states. Note that the 
minimum of $\sigma^2$ would correspond to $n_{p}=n_{d}$ which further 
motivates the definition of Eq. (\ref{eq:sigma}). 
We can also probe the ZR character of holes forming the CDW by 
looking at the magnetization of holes in the rungs, cf. Fig. 
\ref{fig:den}(e). The magnetization $m_{d}$ grows with 
increasing $U$ and for large $U\sim 12t$ it is still around 30\% smaller 
than for localized ZR singlets. However, even in this 
range of $U$ the magnetization on the O sites $m_{p}$ is quite 
small and much below the value for ideal ZR singlets (around 70\% smaller).  
This confirms that the subtle nature of the ZR singlets can be 
only partly captured within the HF approach. 

Remarkably, changing the value of $\Delta$ for fixed $U=8t$ does not 
destabilize the CDW [Fig. \ref{fig:den}(b)] irrespectively of the period. 
This suggests that the charge order is triggered by the on-site Coulomb 
repulsion. However, the character of the holes forming the CDW changes 
and $\sigma^2$ is small ($\sigma^2\sim 0.07$) only as long as $\Delta$ 
is large [Fig. \ref{fig:den}(d)]. 
This is also visible in Fig. \ref{fig:den}(f) where a similar 
discussion as the one concerning Fig. \ref{fig:den}(e) applies. 

\begin{figure}[t]
\includegraphics[width=7.5cm]{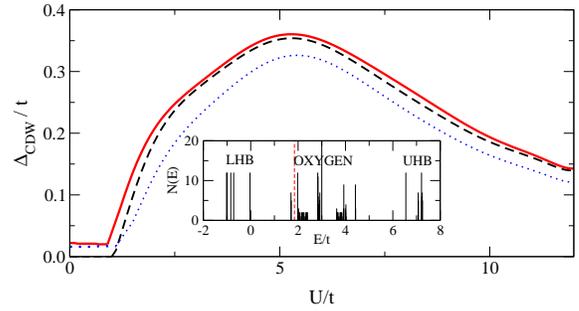}
\caption{(Color online)
Charge gap $\Delta_{\rm CDW}$ in the CDW ground state obtained for 
increasing $U$, with period $\lambda=5,4,3$ (solid, dashed, and dotted 
line). The inset shows the density of states $N(E)$ for $U=8t$ and 
$\lambda=5$, with small CDW gap near the Fermi energy 
(dashed line). Parameters: $\Delta=3t$, $U_p=0$.
}
\label{fig:gap}
\end{figure}

To gain a deeper understanding of the results we calculated the charge 
gap $\Delta_{\rm CDW}$ as a function of the Hubbard $U$, see Fig. 
\ref{fig:gap}. One finds that the gap decreases with the CDW period
which explains the behavior of $p$ observed for different 
periods [Fig. \ref{fig:den}(a)]. In general the dependence of 
$\Delta_{\rm CDW}$ on $U$ qualitatively mimics the relation between 
$p$ and $U$ which suggests that the CDW gains stability when
an insulating state is formed. Indeed,
the electronic density of states $N(E)$ (inset of Fig. \ref{fig:gap})
shows well developed lower and upper Hubbard bands (LHB and UHB) 
separated by an oxygen band, with a small CDW gap in the latter band.
Altogether, one finds that:
(i) the Coulomb interaction $U$ can stabilize the CDW in the Cu$_2$O$_5$ 
ladders, (ii) the CDW phase can be viewed as an equidistant distribution 
of the ZR singlet states in the relevant parameter regime, and
(iii) all of the stable periods (even and odd) behave similarly. 

Next, we investigate the influence of the interladder coupling. 
At finite $U_p$ the "external" fields 
$\rho_{j\alpha}=\langle n_{j+\lambda,\alpha}\rangle$ in Eq. 
(\ref{eq:one}) contribute and 
were self-consistently determined by iterating the HF equations. Thereby, 
the symmetry of the CDW state was chosen in such a way that the rungs 
were translated by $\lambda$ Cu-O lattice constants ($\lambda$ odd) in 
the neighboring ladders to maximize the distance between them 
(Fig. \ref{fig:art}), which minimizes the HF energy. For even $\lambda=4$ 
the numerical calculations performed with the realistic parameters
\cite{Nish02} for Cu$_2$O$_5$ ladder ($U=8t$ and $\Delta=3t$) 
confirmed that two states shown by dotted ovals in Fig. \ref{fig:art} 
are degenerate, as expected.
The effect of the interladder interaction $U_p$ was identified by 
comparing the ground states derived separately in two cases: 
($A$) with $\rho_{j\alpha}=0$, i.e., using only the (intraorbital) 
      repulsion between oxygen holes on the considered ladder; 
($B$) by implementing the "external" fields $\{\rho_{j\alpha}\}$ 
      calculated self-consistently, i.e., including both the 
      intraorbital and interorbital Coulomb repulsion between 
      holes on oxygen sites.

\begin{figure}[t!]
\includegraphics[width=8.4cm]{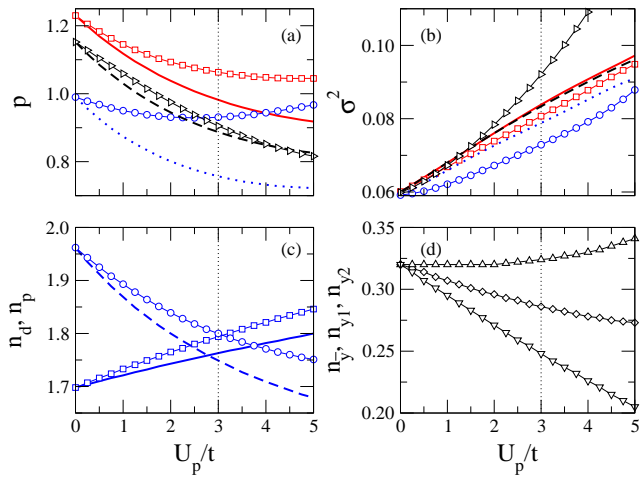}
\caption{(Color online)
The CDW ground state for increasing $U_p$: 
(a) CDW order parameter $p$ and
(b) ZR singlet dispersion $\sigma^2$, 
 for $\lambda=5,4,3$ shown by solid, dashed, 
 and dotted lines (squares, triangles, and 
 circles) in case $A$ ($B$), see text; 
(c) for $\lambda=3$ charge on Cu (O) sites in the rung  
  shown by solid (dashed) line in case $A$ and by squares 
 (circles) in case $B$, see Eq. (\ref{eq:n});
(d) for $\lambda=4$ charge in different $y$ orbitals 
 ($n_{\bar{y}}$, $n_{y1}$, and $n_{y2}$, shown by diamonds, triangles
down, and up) in the rung in case ($B$), see text. 
Vertical lines mark the realistic value (Ref. \onlinecite{Nish02}) of $U_p=3t$.
Parameters: $\Delta=3t$, $U=8t$.   
}
\label{fig:up}
\end{figure}

One finds that in case $A$ the CDW order parameter $p$ decreases in 
a similar way for all periods, cf. Fig. \ref{fig:up}(a),
as well as for even period ($\lambda=4$) when the interladder coupling is 
switched on (case $B$). Remarkably, a qualitatively distinct behavior 
is found for {\it odd\/} periods -- here the interladder coupling 
supports the onset of the CDW phase and the order parameter 
either saturates or even increases with increasing strength of the 
on-site repulsion $U_p$ (as for $\lambda=3$), see also Fig. \ref{fig:up}(c). 
In fact, the interladder coupling enhances the hole density in the rungs.

Another striking effect is the qualitatively distinct behavior of the 
ZR dispersion $\sigma^2$ for odd and even periods, see Fig. 
\ref{fig:up}(b). While for period $\lambda=4$ switching on the 
interladder coupling ($B$) drastically increases $\sigma^2$ with 
respect to the single ladder case ($A$), the results are precisely 
opposite for odd periods $\lambda=3,5$. Furthermore, this increase of 
$\sigma^2$ with $U_p$ in case ($B$) is large for even period -- 
its value $\sim 0.1$ found for large (but still realistic) $U_p\sim 3.5t$ 
is comparable to the value of the ZR dispersion for a single ladder with 
$\Delta\sim t$ [Fig. \ref{fig:den}(d)], where we do not expect stable 
ZR singlets. This large increase of $\sigma^2$ in this case follows from 
the geometrical frustration of the CDW state, as for even periods the two 
$y$ orbitals in the same rung are not equivalent [one of them (say $y1$) 
is closer than the other one (say $y2$) to the rung in the neighboring 
ladder], as shown in Fig. \ref{fig:up}(d). We have also verified that
the mean hole density 
$n_{\bar{y}}=\frac{1}{2}(n_{y1}+n_{y2})$ almost does not change 
when the interladder coupling is switched off (not shown).

Thus, we conclude that the interladder interaction: 
(i) supports the CDW states with odd periods $\lambda=3,5$ and slightly 
    disfavors the frustrated CDW state with even period $\lambda=4$,
(ii) destabilizes the homogeneous ZR-like distribution of holes in the 
     rungs for period $\lambda=4$. 
In contrast, experimentally one finds that in SCCO with $x=4$ 
($n_h\sim 1.25$) the holes are distributed isotropically over O sites 
in the rung,\cite{Rus06b} but the CDW is unstable.\cite{Rus06a} 
We suggest that, since in reality the ZR singlets are much more rigid 
than the present classical ZR states (as the energy gain due to quantum 
fluctuations and phase coherence are not captured in these states) and 
in reality the system is less prone to order than in the HF approximation, 
the interladder interactions in the model Eq. (\ref{eq:one}) would 
indeed destabilize the CDW with even period. 

In summary, we have shown that the CDW combined with the SDW can be 
stabilized in the spin ladders of SCCO merely due to on-site Coulomb 
repulsion on Cu sites. 
The presented results explain the experimentally observed CDW states with 
odd periods for $x=0$ and $x=11$, and provide a theoretical explanation 
why the CDW states with even period could not be observed.\cite{Rus06a} 
In addition our results suggest that an extension to the two leg ladder 
$t$-$J$ model used in Ref. \onlinecite{Whi02} is needed to capture the 
subtle properties of the CDW states in SCCO. The simplest extension might 
be to consider a pair of ladders in a $t$-$J$ model plus an interladder 
Coulomb repulsion to represent the physics described in our study. 
We are looking forward to future studies of this kind 
using the much more sophisticated DMRG-like methods.
 
This work was supported by the Polish Ministry of Science and Education 
under Project No. N202~068~32/1481 and by the Canadian research funding 
organizations NSERC, CFI, and CRC.

{\it Note added in proof\/}: In contrast to the present charge
transfer model, Roux {\it et al.} \cite{Rouun} have found that the CDW state
is unstable for other hole densities than $n_h=1.25$ and $n_h=1.50$ in
their recent DMRG and bosonization studies of the $t$-$J$ model for
the single ladder with $1.0 < n_h \leq 1.5$.


\end{document}